\documentclass[prc,preprint,superscriptaddress,showpacs,nofootinbib,%
tightenlines]{revtex4} 
\usepackage{graphics} 
\usepackage{epsfig} 
\def\nn{\nonumber} 
\def\gcirc{\mbox{\it \r{g}}}

\def\Dslash{\hspace{-2.4mm}\slash\hspace{1.4mm}} 
\def\gA0{\stackrel{\circ}{g_A}} 
\begin{document} 
\preprint{MKPH-T-03-7} 
\title{Electromagnetic form factors of the nucleon in relativistic 
baryon chiral perturbation theory} 
\author{Thomas Fuchs} 
\affiliation{Institut f\"ur Kernphysik, Johannes 
Gutenberg-Universit\"at, D-55099 Mainz, Germany} 
\author{Jambul Gegelia} 
\thanks{Alexander von Humboldt Research Fellow} 
\affiliation{Institut f\"ur Kernphysik, Johannes 
Gutenberg-Universit\"at, D-55099 Mainz, Germany} 
\affiliation{High Energy Physics Institute, 
Tbilisi State University, 
University St.~9, 380086 Tbilisi, Georgia} 
\author{Stefan Scherer} 
\affiliation{Institut f\"ur Kernphysik, Johannes 
Gutenberg-Universit\"at, D-55099 Mainz, Germany} 
\date{May 23, 2003} 
\begin{abstract} 
   The electromagnetic form factors of the nucleon are calculated at 
order $q^4$ in the relativistic formulation of baryon chiral 
perturbation theory.  
   In order to obtain a consistent power counting for the renormalized  
diagrams we make use of the extended on-mass-shell renormalization 
scheme which we have discussed in a recent paper.  
   We analyze the Dirac and Pauli form factors, $F_1$ and $F_2$, as well as 
the electric and magnetic Sachs form factors, $G_E$ and $G_M$.  
   Our results are compared with those obtained in the heavy-baryon 
approach and in the (relativistic) infrared regularization.  
   In addition, the Sachs form factors are compared with  
experimental data.  
   We confirm previous findings that a one-loop calculation 
of the electromagnetic form factors does not generate sufficient 
curvature for the Sachs form factors $G^p_E$, $G^p_M$, and 
$G^n_M$. 
   Moreover, the electric form factor of the neutron is very sensitive 
to higher-order contributions.  
\end{abstract} 
\maketitle 
\section{Introduction} 
   Chiral perturbation theory ($\chi$PT) for mesons  
\cite{Weinberg:1978kz,Gasser:1984yg} is the effective field theory of  
low-energy QCD in the vacuum sector (baryon number zero),  
expressed in terms of the effective degrees of freedom active at low energies. 
   These are the Goldstone bosons resulting from the  
spontaneous symmetry breaking of the chiral group  
$\mbox{SU}(N)_L\times\mbox{SU}(N)_R$, reflecting  
the symmetry of the QCD Lagrangian for $N$ massless quark flavors,  
down to its vectorial subgroup $\mbox{SU}(N)_V$, denoting the symmetry of  
the vacuum.  
   The effective Lagrangian is organized in terms of a simultaneous expansion  
in powers of (covariant) derivatives and quark-mass terms, and a perturbative 
expansion of observables is expected to work for momenta sufficiently small 
in comparison with the scale of chiral symmetry breaking, 
$\Lambda_\chi \sim 1 \ \mbox{GeV}$. 
   Perturbative calculations in the mesonic sector are feasible due 
to the existence of a consistent power counting scheme  
\cite{Weinberg:1978kz,Gasser:1984yg} establishing a connection between the 
chiral expansion and the loop expansion 
(see, e.g., Ref.\ \cite{Scherer:2002tk} for a recent overview). 
 
   The one-nucleon sector (baryon number one) turns out to be more  
complicated \cite{Gasser:1988rb} 
because of the presence of a new mass scale---the mass of the  
nucleon---which does not vanish in the chiral limit.  
   In particular, loop diagrams containing internal nucleon lines may  
contribute to low orders in the chiral expansion.  
   Therefore, the relation between the chiral expansion and the loop  
expansion appears to be lost \cite{Gasser:1988rb}.  
   This problem has been solved in the framework of heavy-baryon  
chiral perturbation theory (HB$\chi$PT)  
\cite{Jenkins:1990jv,Bernard:1992qa}, 
where the power counting is restored in terms of an additional  
$1/m$ expansion, and perturbative calculations are again possible.  
   However, HB$\chi$PT has its own shortcomings: the corresponding  
perturbation series fails to converge in part of the low-energy region 
\cite{Bernard:1996cc}.  
   Moreover, at higher orders in the chiral expansion, the expressions 
due to $1/m$ corrections of the Lagrangian become increasingly complicated 
\cite{Ecker:1995rk,Fettes:2000gb}. 
   These disadvantages are related to the nonrelativistic momentum expansion  
in this approach. 
 
   Recently, various methods have been proposed to establish a consistent 
power counting within a relativistic approach 
\cite{Tang:1996ca,Ellis:1997kc,Becher:1999he,Gegelia:1999gf,Gegelia:1999qt,%
Lutz:1999yr,Lutz:2001yb,Fuchs:2003qc}. 
   In this work, we will discuss the electromagnetic form factors of the  
nucleon within the extended on-mass-shell (EOMS) renormalization scheme  
of Ref.\ \cite{Fuchs:2003qc}. 
   In this scheme renormalization is performed by expanding a 
given Feynman diagram in terms of small quantities and  
subtracting those terms which violate the power counting.  
   Since the subtracted terms are regular in the small quantities, they 
can be absorbed in a finite number of low-energy constants of the chiral  
Lagrangian.  
   So far, the new method has been applied to a discussion of the mass 
and the scalar form factor of the nucleon at ${\cal O}(q^4)$ 
\cite{Fuchs:2003qc}. 
 
   The electromagnetic form factors parameterize the single-nucleon matrix  
element of the electromagnetic current operator and thus reflect the 
electromagnetic structure of the nucleon. 
   The matrix element is usually described in terms of either the 
Dirac and Pauli form factors, $F_1$ and $F_2$, or the electric and 
magnetic Sachs form factors, $G_E$ and $G_M$. 
   The latter are popular, because their Fourier transforms in the 
Breit frame come closest to an interpretation as the distribution 
of charge and magnetization inside the nucleon \cite{ESW 60,Sac 62} 
(for a recent discussion and an overview of the existing relevant 
form factor measurements, see Refs.\ \cite{Kelly:2002if,Friedrich:2003iz}). 
   For low momentum transfers, $Q^2 \le 1$ GeV$^2$, the Sachs form factors  
$G_E^p$, $G_M^p$, and $G_M^n$ are reasonably well described  
by a dipole form factor $G_D$,  
$$ 
  G_E^p \approx \frac{G_M^p}{\mu_p} \approx \frac{G_M^n}{\mu_n} \approx G_D, 
$$ 
where $G_D(Q^2)= [1 + Q^2/(0.71 \,\mbox{GeV}^2)]^{-2}$. 
   As has been stressed in Ref.\ \cite{Friedrich:2003iz}, there is  
essentially no physical motivation for such a form and it is also not able  
to describe ``fine structure'' in the data. 
   The electric form factor of the neutron, $G_E^n$, is the least 
precisely known due to the lack of a free neutron target and the 
fact that it is small. 
   However, recent polarization experiments have improved our empirical 
knowledge of $G_E^n$ considerably (see Ref.\ \cite{Friedrich:2003iz} 
and references therein). 
  Clearly, a quantitative description of these four form factors  
is a stringent test for any theory or model of the strong interactions
(see, e.g., Refs.\ 
\cite{Iachello:1972nu,Kroll:1990hg,Ji:ff,Schlumpf:1992pp,Cardarelli:1995dc,%
Christov:1995hr,Mergell:1995bf,Lu:1997sd,Boffi:2001zb}).  
   In chiral perturbation theory, these form factors have been 
calculated within the early relativistic approach \cite{Gasser:1988rb}, 
HB$\chi$PT \cite{Bernard:1992qa,Fearing:1997dp}, 
the small-scale expansion \cite{Bernard:1998gv}, 
and within the relativistic infrared regularization approach  
\cite{Kubis:2000zd}. 
 
   Our work is organized as follows.  
   In Sec.\ \ref{chiral_lagrangians} we shortly review the effective  
Lagrangian and list the interaction terms relevant for our calculation. 
   In Sec.\ \ref{emff} the Dirac and Pauli as well as the Sachs 
form factors are discussed.  
   Our results are compared with the HB$\chi$PT calculation, the method of  
infrared regularization, and experimental data.  
   A summary and some concluding remarks are given in Sec.\ \ref{summary}.
   Some technical details can be found in the appendices. 
 
\section{\label{chiral_lagrangians} Chiral Lagrangian} 
   The relevant effective Lagrangian contains two pieces,  
$$  
{\cal L}_{\rm eff} = {\cal L}_{\pi} + {\cal L}_{\pi N},  
$$  
where the first part is the pure mesonic Lagrangian and the second part 
consists of terms which are bilinear in the isospin doublet $\Psi$ 
containing the proton and neutron fields.  
   The different terms of the effective Lagrangian can be organized according  
to the chiral derivative and quark-mass expansion  
\cite{Weinberg:1978kz,Gasser:1984yg,Gasser:1988rb}, 
\begin{equation} 
   {\cal L}_{\rm eff} = {\cal L}_{2} + {\cal L}^{(1)}_{\pi N} + 
   {\cal L}^{(2)}_{\pi N} + {\cal L}^{(3)}_{\pi N} + {\cal L}^{(4)}_{\pi N} 
       + \cdots, 
\end{equation} 
where the subscripts (superscripts) in ${\cal L}_\pi$ (${\cal L}_{\pi N}$) 
refer to the order in the expansion, and the ellipsis denotes terms of higher 
order which are not relevant for our calculation. 
 
   From the mesonic sector we only need the lowest-order Lagrangian 
[${\cal O}(q^2)$] including the quark-mass term  
and the coupling to an external electromagnetic field ${\cal A}^\mu$ 
in terms of the isovector vector field  
$v^\mu=-e\frac{\tau_3}{2}{\cal A}^\mu$ ($e^2/4\pi\approx 1/137$, $e>0$) 
\cite{Gasser:1984yg},\footnote{In the 
mesonic sector, the isoscalar coupling to the external electromagnetic 
field only becomes relevant at ${\cal O}(q^6)$  
\cite{Ebertshauser:2001nj}.} 
\begin{eqnarray} 
{\cal L}_2 &=& \frac{F^2}{4}\mbox{Tr}\left(\partial_\mu U \partial^\mu 
U^\dagger \right) 
+\frac{F^2  M^2}{4}\mbox{Tr}\left(U^\dagger+ U\right) 
+ i\frac{F^2}{2}\mbox{Tr}\left[ (\partial_\mu U U^\dagger + \partial_\mu 
U^\dagger U) v^\mu\right]+\cdots. 
\end{eqnarray} 
   Here, $F$ denotes the pion decay constant in the chiral limit, 
$F_\pi = F[1+{\cal O}(\hat{m})] = 92.4$ MeV, and  
$M^2=2 B\hat{m}$ 
is the lowest-order prediction for the pion mass squared, where $B$ is  
related to the quark condensate $\langle\bar{q}q\rangle_0$ in the chiral 
limit \cite{Gasser:1984yg,Colangelo:2001sp}.  
   We make use of the isospin-symmetric limit $m_u=m_d=\hat{m}$. 
   The triplet of pion fields is collected in the matrix $U$,  
\begin{eqnarray*} 
   U(x) &=& u^2(x) = \exp\left(\frac{i \Phi(x)}{F}\right), \\ 
   \Phi &=& \vec{\tau}\cdot \vec{\phi} = \left( 
         \begin{array}{cc} 
            \pi^0 & \sqrt{2}\pi^+ \\ \sqrt{2}\pi^- & -\pi^0 
         \end{array}\right). 
\end{eqnarray*} 
 
   The lowest-order pion-nucleon Lagrangian is given by 
\cite{Gasser:1988rb} 
\begin{equation} 
\label{nucl1} 
  {\cal L}^{(1)}_{\pi N}=\bar{\Psi}\left(i D\Dslash - m 
     + \frac{\gcirc_A}{2} \gamma^\mu\gamma_5 u_\mu \right) \Psi, 
\end{equation} 
with 
\begin{eqnarray*} 
   D_\mu\Psi &=&\left(\partial_\mu +\Gamma_\mu -i v_\mu^{(s)} 
   \right)\Psi,\nn\\ 
 \Gamma_\mu &=& \frac{1}{2}\left[u^\dagger\partial_\mu u 
+u\partial_\mu u^\dagger-i(u^\dagger v_\mu u + u v_\mu u^\dagger)\right],\nn\\ 
   u_\mu &=& i\left[u^\dagger\partial_\mu u - u\partial_\mu u^\dagger 
   -i(u^\dagger v_\mu u - u v_\mu u^\dagger)\right]. 
\end{eqnarray*} 
   In Eq.\ (\ref{nucl1}), $m$ and  
$\gcirc_A$ refer to the chiral limit (at fixed strange-quark mass) of the  
physical nucleon mass and the axial-vector coupling constant, respectively. 
   In the definition of the covariant derivative we follow  
Ref.\ \cite{Ecker:1995rk}, where $\Gamma_\mu$ only contains traceless  
external fields and the coupling to the isosinglet vector field  
$v_\mu^{(s)}$ is considered separately.
 
   The next-to-leading-order pion-nucleon Lagrangian contains seven 
low-energy constants $c_i$ \cite{Gasser:1988rb,Fettes:2000gb}, where only 
the terms proportional to $c_1$, $c_2$, $c_4$, $c_6$, and $c_7$ contribute 
to the calculation of the electromagnetic form factors at ${\cal O}(q^4)$, 
\begin{eqnarray} 
\label{nucl2} 
   {\cal L}_{\pi N}^{(2)} &=& 
   c_1 \mbox{Tr}(\chi_{+})\bar\Psi\Psi
    - \frac{c_2}{4m^2}\left[ \bar{\Psi}\mbox{Tr}\left(u_\mu u_\nu\right) 
      D^\mu D^\nu \Psi + \mbox{h.c.}\right] 
\nonumber\\&& 
+ \bar{\Psi}\left[ i\frac{c_4}{4} \left[u_\mu, u_\nu\right] 
      + \frac{c_6}{2}f_{\mu\nu}^+ 
    + \frac{c_7}{2} v_{\mu\nu}^{(s)}\right] \sigma^{\mu\nu} \Psi +\cdots, 
\nonumber\\ 
\end{eqnarray} 
where h.c.~refers to the Hermitian conjugate and 
\begin{eqnarray*} 
\chi_+& = &M^2(U+U^\dagger),\nonumber\\
v_{\mu\nu}^{(s)} &=& \partial_\mu v_\nu^{(s)} - \partial_\nu v_\mu^{(s)},\nn\\ 
 f_{\mu\nu}^\pm &=& u f_{\mu\nu}^L u^\dagger \pm u^\dagger f_{\mu\nu}^Ru,\nn\\ 
 f_{\mu\nu}^L &=& \partial_{\mu}l_{\nu} - \partial_{\nu}l_{\mu} - 
                    i\left[l_{\mu},l_{\nu}\right],\nn\\ 
 f_{\mu\nu}^R &=& \partial_{\mu}r_{\nu} - \partial_{\nu}r_{\mu} - 
                    i\left[r_{\mu},r_{\nu}\right], 
\end{eqnarray*} 
   with $r_\mu=l_\mu=-e\tau_3{\cal A}_\mu/2$
and $v_\mu^{(s)}=-e {\cal A}_\mu/2$. 
   With the convention of choosing $f_{\mu\nu}^+$ to be traceless,  
$c_6$ ($c_7$) will be related to the isovector (isoscalar) magnetic moment 
of the nucleon in the chiral limit. 
 
 The complete Lagrangians at third and fourth order can be found in Refs.\ 
\cite{Ecker:1995rk,Fettes:2000gb} and \cite{Fettes:2000gb}, respectively.  
   Again, we only list the terms needed for the calculation of the  
electromagnetic form factors,\footnote{We took the Lagrangians 
of Ref.\ \cite{Fettes:2000gb} and made the replacements  
$\tilde{F}^+_{\mu\nu}\to f^+_{\mu\nu}$ and 
$\mbox{Tr}(F^+_{\mu\nu})\to 4 v^{(s)}_{\mu\nu}$.} 
\begin{eqnarray*} 
    {\cal L}_{\pi N}^{(3)} &=& \frac{i}{2m} 
      d_6\bar{\Psi}\left[ D^\mu,f_{\mu\nu}^+ \right] 
     D^\nu \Psi + \mbox{h.c.}\nonumber\\ 
     && + \frac{2i}{m} d_7 \bar{\Psi}(\partial^\mu 
      v_{\mu\nu}^{(s)}) D^\nu \Psi + \mbox{h.c.}+\cdots,  \nn\\ 
    {\cal L}_{\pi N}^{(4)} &=& \bar{\Psi}\left[- 2\,e_{54}\, 
     \left( \partial^\lambda \partial_\lambda v_{\mu\nu}^{(s)} \right) 
     - \frac{1}{2}\,e_{74}\, [D^\lambda,[D_\lambda, 
     f_{\mu\nu}^+]]  \right. \nn\\ && \left. 
     - 2 \,e_{105}\, v_{\mu\nu}^{(s)} \mbox{Tr}\left(\chi_+\right) 
      - \frac{1}{2}\,e_{106}\, f_{\mu\nu}^+ 
     \mbox{Tr}\left( \chi_+\right) \right]\sigma^{\mu\nu} \Psi + \cdots. 
\end{eqnarray*} 
   The values for the low-energy constants (LECs) have to be determined from  
empirical input.  
   In general, their numerical values depend on the renormalization scheme 
in question. 
   As we will see later, some of the constants can be fitted to the anomalous 
magnetic moments and to the charge and magnetic radii of the nucleon. 
 
\section{\label{emff}Electromagnetic form factors of the nucleon} 
\subsection{Definition of the electromagnetic form factors} 
 
   In terms of QCD degrees of freedom the interaction with an external 
electromagnetic field ${\cal A}_\mu$ is given by 
$$ 
{\cal L}_{\rm e.m.}=-e J^\mu {\cal A}_\mu, 
$$ 
where the electromagnetic current operator reads
\begin{equation} 
\label{jmudef} 
  J^\mu(x) = \frac{2}{3}\bar{u}(x)\gamma^\mu u(x) - \frac{1}{3}\bar{d}(x) 
    \gamma^\mu d(x) +\cdots 
    = \bar{q}(x) \left(\frac{1}{6} + \frac{\tau_3}{2}\right)\gamma^\mu q(x) 
+\cdots. 
\end{equation} 
   The ellipsis denotes the contribution due to the heavier quarks which
we do not consider here.
   The electromagnetic form factors are defined via the matrix  
element 
\begin{equation} 
\label{emffdef} 
   \langle N(p_f)\left| J^\mu(0) \right| N(p_i) \rangle = 
     \bar{u}(p_f)\left[ \gamma^\mu F_1^N(Q^2) + \frac{i\sigma^{\mu\nu}q_\nu} 
     {2m_N}F_2^N(Q^2) \right] u(p_i), \qquad N=p,n, 
\end{equation} 
where $q=p_f-p_i$ is the momentum transfer and 
$Q^2\equiv-q^2=-t \ge 0$.\footnote{Since we discuss the form factors in the  
space-like region, here we adopt the convention of taking $Q^2=-t$ as the  
argument of the form factors as is common practice in the context of electron  
scattering.}  
   The functions $F^N_1(Q^2)$ and $F^N_2(Q^2)$ are the Dirac and Pauli 
form factors of the nucleon, respectively.  
   At $Q^2=0$, these form factors are given by the electric charges and  
the anomalous magnetic moments in units of the charge and the nuclear  
magneton, respectively: 
\begin{equation} 
\label{magnmom} 
   F_1^p(0) = 1, \qquad F_1^n(0) = 0, \qquad 
   F_2^p(0) = \kappa_p=1.793, \qquad F_2^n(0) = \kappa_n=-1.913. 
\end{equation} 
   In that sense, Eq.\ (\ref{emffdef}) provides a ``natural'' extension 
of the electromagnetic vertex of a ``point particle'' with an anomalous 
magnetic moment.\footnote{For a discussion of the implications of gauge  
invariance in its strong form and alternative forms of the vertex, 
see, e.g., Ref.\ \cite{Scherer:1996ux}.} 
   In the actual calculation, it is more convenient to work in the 
isospin basis 
\begin{equation} 
  F_i^{(s)} = F_i^p + F_i^n,\qquad 
  F_i^{(v)} = F_i^p - F_i^n, \qquad i=1,2, 
\end{equation} 
so that the electromagnetic form factors are obtained as follows, 
$$ 
    F_i^N = \frac{1}{2} F_i^{(s)} + \frac{\tau_3}{2} F_i^{(v)},\qquad i=1,2. 
$$ 
    
   Experimental data are usually analyzed in terms of the  
electric and magnetic Sachs form factors \cite{ESW 60,Sac 62},  
$G_E^N(Q^2)$ and $G_M^N(Q^2)$, defined by 
\begin{eqnarray} 
G^N_E(Q^2) &=& F^N_1(Q^2) - \frac{Q^2}{4m_N^2}F_2^N(Q^2),\label{GEN}\\ 
G^N_M(Q^2) &=& F^N_1(Q^2) + F_2^N(Q^2).\label{GMN} 
\end{eqnarray} 
   The Fourier transforms of the Sachs form factors in the Breit frame  
can be related to the distribution of charge and magnetization inside the  
nucleon. 
   For a recent discussion of the limits of such an interpretation, 
see Ref.\ \cite{Kelly:2002if}.

\subsection{Calculation of the Dirac and Pauli form factors} 
 
   The graphs contributing to the electromagnetic current matrix element of  
Eq.\ (\ref{emffdef}) up to and including ${\cal O}(q^4)$ are shown  
in Fig.\ \ref{emffgraphen} (we do not display external leg 
corrections).\footnote{In order to take care of the mass shift due to 
${\cal L}^{(2)}_{\pi N}$, we use $m_2=m-4 c_1 M^2$ in the
fermion propagators \cite{Becher:1999he}. 
   At the end of the calculation we then replace $m_2$ by the physical
nucleon mass $m_N$, because the difference shows up only in higher orders.}  
   In particular, the loop graphs labeled by the numbers (5) to (9)  
in Fig.\ \ref{emffgraphen} contribute at ${\cal O}(q^3)$ while the graphs  
(10) to (12) contribute at ${\cal O}(q^4)$.  
   The individual unrenormalized contributions to the form factors $F_1$ and  
$F_2$ can be found in App.\ \ref{emffanh}. 
   To renormalize the form factor diagrams we first apply the 
subtraction scheme used by Gasser and Leutwyler 
\cite{Gasser:1984yg,Gasser:1988rb} which we denote by modified 
minimal subtraction scheme of $\chi$PT ($\widetilde{\rm MS}$).  
   We then perform the additional finite subtractions according to 
our EOMS scheme (see Ref.\ \cite{Fuchs:2003qc} for details).  
   For this we determine the chiral order of a given diagram by applying  
the standard power counting and then subtract those terms which violate the  
power counting. 
   Such finite subtractions are achieved by expanding the quantities of the 
$\widetilde{\rm MS}$ scheme in terms of the parameters of our EOMS scheme.  
   This expansion produces the counterterm contributions 
which are responsible for the required additional subtractions.  
   In view of Eq.\ (\ref{emffdef}), an ${\cal O}(q^4)$ calculation  
of the electromagnetic form factors yields $F_1$ to ${\cal 
O}(q^3)$ and $F_2$ to ${\cal O}(q^2)$, since both the polarization 
vector $\epsilon^\mu$ and the four-momentum $q^\mu$ of the virtual photon  
count as ${\cal O}(q)$.  
   Diagrams potentially violating the power counting are loop diagrams 
with internal nucleon lines.  
   In the present case, we find that the diagrams (5), (8), and (10) of  
Fig.\ \ref{emffgraphen} need subtractions beyond $\widetilde{\rm MS}$
whereas this is not necessary 
for the diagrams (7).  
   The subtraction term for the Dirac form factor 
reads
$$ 
\Delta F_1^{\,10} =  
\frac{\gA0^2m}{64\pi^2F^2}\left(3c_7 - 2c_6 \tau_3\right)t, 
$$ 
and similarly for the Pauli form factor\footnote{One factor of $m_N$ is
due to the fact that the parameterization of Eq.\ (\ref{emffdef}) 
contains an explicit factor $1/m_N$ multiplying the Pauli form factor.}
\begin{eqnarray*} 
\Delta F_2^{\,5} &=& -\frac{\gA0^2m_N (m-4c_1 M^2)}{32\pi^2F^2}
\left(3-\tau_3\right),\\ 
\Delta F_2^{\,8} &=& \frac{\gA0^2m_N (m-4 c_1 M^2)}{8\pi^2F^2}\tau_3,\\ 
\Delta F_2^{\,10} &=& - \frac{\gA0^2 m_N(m^2-8c_1 M^2 m)}{16\pi^2F^2} 
\left(3c_7 - 2c_6\tau_3 \right). 
\end{eqnarray*} 
   For completeness, we determine the expansion of the couplings of the 
$\widetilde{\rm MS}$ scheme  $c^r_6$, $c^r_7$, $d^r_6$, $d^r_7$, 
$e^r_{54}$, and $e^r_{74}$ in terms of our renormalized parameters,\footnote{
For notational convenience, we omit a subscript R for the EOMS-renormalized
parameters, but one should not confuse them with the {\em bare} parameters
of the bare Lagrangian.} 
\begin{eqnarray*} 
c_6^r &=& c_6 - \frac{\gA0^2m}{64\pi^2F^2}\left(5+4m c_6\right),\nonumber\\ 
c_7^r &=& c_7 + \frac{3\gA0^2m}{32\pi^2F^2}\left(1+2mc_7\right),\nonumber\\ 
d_6^r &=& d_6 - \frac{\gA0^2m}{32\pi^2 F^2}c_6,\nonumber\\ 
d_7^r &=& d_7 + \frac{3\gA0^2m}{128\pi^2 F^2}c_7,\nonumber\\ 
e_{54}^r &=& e_{54} - \frac{3\gA0^2}{256\pi^2 F^2}c_7,\nonumber\\ 
e_{74}^r &=& e_{74} +\frac{\gA0^2}{64\pi^2 F^2}c_6,\\
e_{105}^r&=&e_{105}+\frac{\gA0^2 c_1}{128\pi^2 F^2}(3+12 m c_7),\\
e_{106}^r&=&e_{106}-\frac{\gA0^2 c_1}{64\pi^2 F^2}(5+8m c_6),
\end{eqnarray*} 
producing all the required counterterm contributions. 
 
   In order to obtain the final expression one has to sum up all 
contributions and to multiply the result with the wave function 
renormalization constant $Z$,\footnote{For $F_2$, one needs $Z$ 
only up to order ${\cal O}(q^2)$ due to the factor $q_\nu$ in Eq.\ 
(\ref{emffdef}).} given by \cite{Fuchs:2003qc} 
\begin{equation} 
\label{ZNgeg} 
   Z = 1 - \frac{9\stackrel{\circ}{g_A}^2 M^2}{32\pi^2F^2}\left[ 
       \frac{2}{3} + \ln\left(\frac{M}{m}\right)\right] 
       + \frac{9\stackrel{\circ}{g_A}^2 M^3}{64\pi F^2 m} + O(M^4). 
\end{equation} 
 
   Next, we need to fix the parameters in order to get a graphical  
representation of the form factors.  
   The values of the renormalized LECs $c_2$ and $c_4$ are taken from a 
(tree-level) fit \cite{Becher:2001hv} to the $\pi N$ scattering threshold  
parameters of Ref.\ \cite{Koch:bn}:  
$c_2 = 2.5m_N^{-1}$ and $c_4 = 2.3m_N^{-1}$. 
   Since the $c_i$ only enter loop expressions at ${\cal O}(q^4)$ a 
determination at ${\cal O}(q^2)$ is consistent with the present accuracy. 
    
   The chiral expansion of the anomalous isoscalar and isovector magnetic 
moments can be written as  
\begin{equation} 
\label{kappaexp} 
\kappa^{(s/v)}=\kappa_0^{(s/v)}+\kappa_1^{(s/v)}M 
+\kappa_2^{(s/v)}M^2\ln\left(\frac{M}{m}\right) 
+\kappa_3^{(s/v)}M^2+O(M^3), 
\end{equation} 
where the coefficients, in terms of the EOMS-renormalized parameters, are 
given by 
\begin{eqnarray} 
\label{coefficientskappas} 
\kappa_0^{(s)}&=& 2c_7 m,\nonumber\\ 
\kappa_1^{(s)}&=&0,\nonumber\\ 
\kappa_2^{(s)}&=&-\frac{3(1+2c_7 m)\gA0^2}{8\pi^2F^2} ,\nonumber\\ 
\kappa_3^{(s)}&=&-\frac{3(2+3c_7 m)\gA0^2}{16\pi^2 F^2} -8c_1 c_7 
-32 e_{105}m,\\\label{coefficientskappav} 
\kappa_0^{(v)}&=& 4 c_6 m,\nonumber\\ 
\kappa_1^{(v)}&=& - \frac{\gA0^2 m}{4 \pi F^2},\nonumber\\ 
\kappa_2^{(v)}&=& \frac{4(c_4-c_6)m-\gA0^2(7+8c_6 m)}{8\pi^2 F^2}, 
\nonumber\\ 
\kappa_3^{(v)}&=& -\frac{\gA0^2(1+5 c_6 m)}{8\pi^2 F^2} -16 c_1 c_6
-16 e_{106} m. 
\end{eqnarray} 
   Comparing with Ref.\ \cite{Kubis:2000zd}  
we see that the lowest-order terms as well as the non-analytic terms 
$\sim M\sim \hat{m}^{1/2}$ and $\sim \ln(M)$ coincide, while the analytic 
$\kappa_3^{(s/v)}$ terms differ, because we use a different renormalization
scheme.  
 
   Since chiral perturbation theory does not {\em predict} the anomalous
magnetic moments, we take the empirical values as obtained from 
Eq.\ (\ref{magnmom}) to fix the Pauli form factors at $Q^2=0$.
   The constants $d_6$, $d_7$, $e_{54}$, and $e_ {74}$ were obtained from the 
charge and magnetic radii of the 
nucleon, which are defined via the Sachs form factors  
[see Eqs.\ (\ref{GEN}) and (\ref{GMN})]: 
\begin{eqnarray*} 
   \langle \left(r^p_E\right)^2 \rangle &=& -\frac{6}{G^p_E(0)}\left. 
     \frac{dG^p_E(Q^2)}{dQ^2}\right|_{Q^2=0}, \qquad 
   \langle \left(r^p_M\right)^2 \rangle = -\frac{6}{G^p_M(0)}\left. 
     \frac{dG^p_M(Q^2)}{dQ^2}\right|_{Q^2=0},\\ 
   \langle \left(r^n_E\right)^2 \rangle &=& -6\left.\frac{dG^n_E(Q^2)}{dQ^2} 
     \right|_{Q^2=0}, \qquad\qquad\;\;\; 
   \langle \left(r^n_M\right)^2 \rangle = -\frac{6}{G^n_M(0)}\left. 
      \frac{dG^n_M(Q^2)}{dQ^2}\right|_{Q^2=0}. 
\end{eqnarray*} 
   For the numerical analysis, we make use of the charge and magnetic radii  
as obtained in the dispersion-theoretical analysis of  
Ref.\ \cite{Mergell:1995bf}, $r^p_E = 0.847\, \mbox{fm}$, 
$r^n_E = -0.113\, \mbox{fm}$, $r^p_M = 0.836\, \mbox{fm}$ and 
$r^n_M = 0.889\, \mbox{fm}$. 
   Within the accuracy of our calculation the constants $c_6$, $c_7$,
$e_{105}$, and $e_{106}$ appear in the combinations
\begin{equation} 
   \tilde{c}_6 = c_6 - 4M^2e_{106},\qquad 
   \tilde{c}_7 = c_7 - 16M^2e_{105}. 
\end{equation} 
   Therefore for the numerical analysis we only need the values
for the combinations $\tilde{c}_{6,7}$.
   The values for the LECs are summarized in Table \ref{lecwerte}. 
   From Table \ref{lecwerte} we see that there is quite a big difference  
in the value for the constant $d_6$ as obtained in the infrared regularization 
and our renormalization scheme, respectively. 
   The constants $d_6$ and $d_7$ yield the leading order of the 
electromagnetic radii and thus determine, together with those loop 
graphs which depend on $Q^2$, the slope of the graphs at the origin. 

\begin{table}[ht] 
\begin{center} 
\begin{tabular}{|c||c|c|c|c|c|c|c|c|} 
  \hline 
     & $c_2$ & $c_4$ & $\tilde{c}_6$ & $\tilde{c}_7$ & 
     $d_6$ & $d_7$ & $e_{54}$ & $e_{74}$\\ 
   \hline\hline 
   EOMS & 2.66 & 2.45 & 1.26 & -0.13 & -0.69 & -0.50 & 0.19 & 1.59\\ 
   IR & 2.66 & 2.45 & 1.26 & -0.18 & 0.54 & -0.73 & 0.25 & 1.93\\ \hline 
\end{tabular} 
\end{center} 
\caption{\label{lecwerte} 
   Values for the relevant low-energy constants in the extended on-mass-shell 
renormalization scheme (EOMS) and in the infrared regularization scheme (IR) 
within an ${\cal O}(q^4)$ calculation of relativistic chiral perturbation 
theory. 
   The LECs $c_i$ are given in units of GeV$^{-1}$, the $d_i$ in units of 
GeV$^{-2}$, and the $e_i$ in units of GeV$^{-3}$.} 
\end{table}

\begin{figure}
\epsfig{file=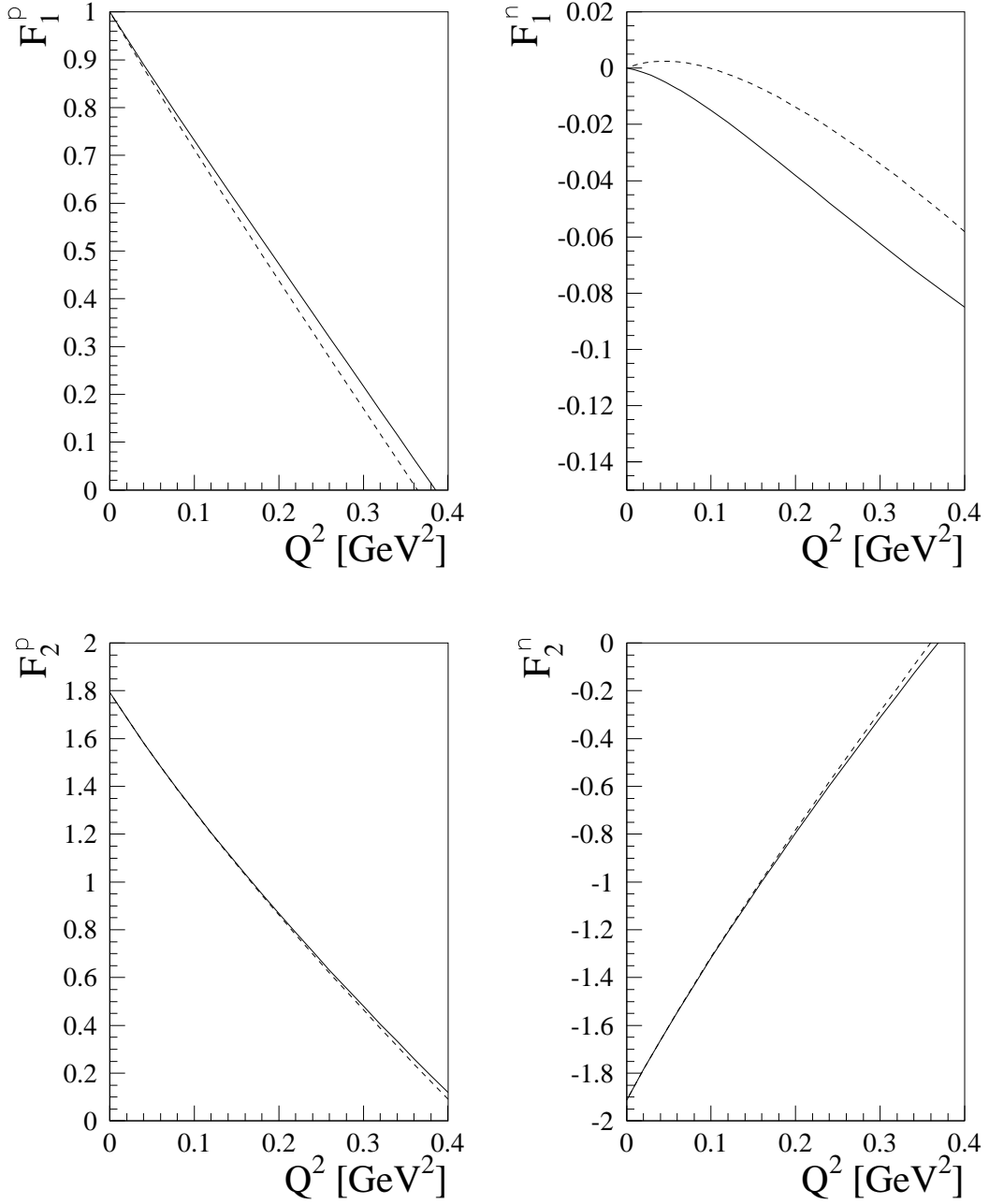, width=14.5truecm} 
\caption{\label{emffnpict}  
The Dirac and Pauli form factors of 
the nucleon in relativistic chiral perturbation theory at ${\cal O}(q^4)$.  
Full lines: our results in the extended on-mass shell (EOMS) scheme;  
dashed lines: infrared regularization result 
\cite{Kubis:2000zd}.} 
\end{figure}

   The results for the Dirac and Pauli form factors are shown 
in Fig.\ \ref{emffnpict} for momentum transfers  
$0\le Q^2 \le 0.4\, \mbox{GeV}^2$.  
   For comparison the results of the 
infrared regularization \cite{Kubis:2000zd} are also shown.  
   We see from Fig.\ \ref{emffnpict} that for the Dirac form factor of the 
proton and for both Pauli form factors the two methods generate 
very similar results. 
   However, this is not the case for the Dirac form factor of the neutron  
which is very small in the entire small $Q^2$ region.  
   We conclude that $F_1^n$ significantly depends on higher-order 
contributions. 
   
\begin{figure} 
\epsfig{file=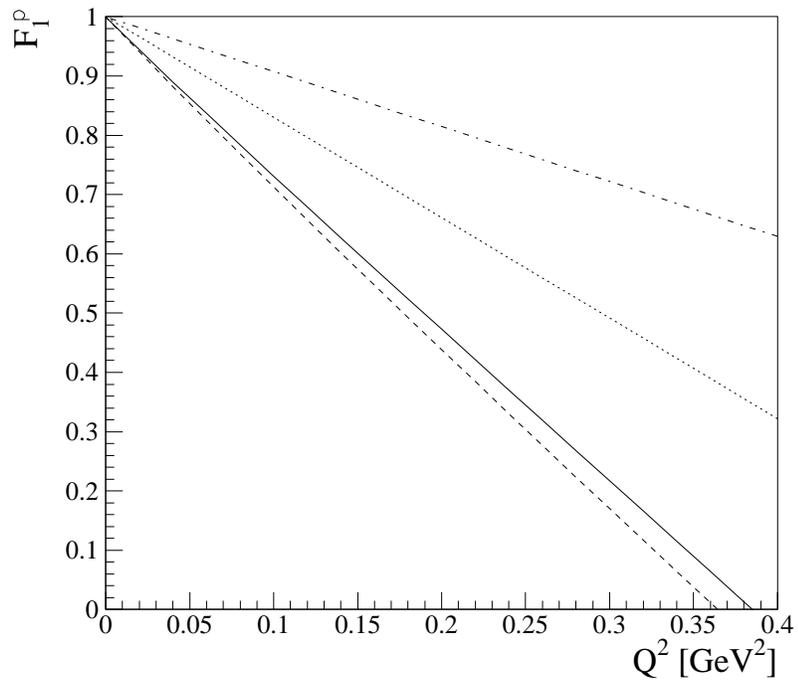, height=9.5truecm} 
\caption{\label{emffnpictvgl} The Dirac form factor of the 
proton at ${\cal O}(q^4)$. 
Solid line: our result; dashed line: infrared regularization; 
dotted line: our result without loop contribution; 
dashed-dotted line: infrared-regularization result without 
loop contribution.} 
\end{figure}

   As an example, let us have a closer look at the Dirac form factor of the 
proton, $F_1^p$, which, for both methods, shows an almost linear behavior 
in $Q^2$, i.e., has very little curvature. 
   In order to show the influence of the loop contributions, in Fig.\ 
\ref{emffnpictvgl} we compare the full results for $F_1^p$ in 
both relativistic renormalization schemes with the contact-graph 
contributions only.
   We see that the loop contributions cannot be neglected, i.e., the 
contribution of the pion cloud plays an important role.  
   Moreover, in the infrared regularization the main contribution to the 
slope is due to loop contributions, while in our renormalization scheme the 
loop contributions are smaller.  
   The situation here is different from the electromagnetic form factor 
of the pion whose linear behavior is predominantly due to a  
contact graph \cite{Gasser:1984yg}.

\subsection{Results for the Sachs form factors} 
   We now turn to a discussion of the Sachs form factors in the EOMS 
renormalization scheme and compare them with experimental data as well as 
the results of the infrared regularization and of HB$\chi$PT
at ${\cal O}(q^3)$ \cite{Bernard:1992qa,Fearing:1997dp}.\footnote{The effects
due to the $\Delta$ resonance, as calculated in the small-scale expansion, 
can be found in Ref.\ \cite{Bernard:1998gv}.} 
   Recall that a full ${\cal O}(q^4)$ calculation yields $G_E$ and 
$G_M$ up to ${\cal O}(q^3)$ and ${\cal O}(q^2)$, respectively.\footnote{
As discussed in Ref.\ \cite{Kubis:2000zd}, when combining $F_1$ and $F_2$ to 
give the Sachs form factors $G_E$ and $G_M$ [see Eqs.\ (\ref{GEN})
and (\ref{GMN})] the chiral orders are mixed.} 

\begin{figure}
\epsfig{file=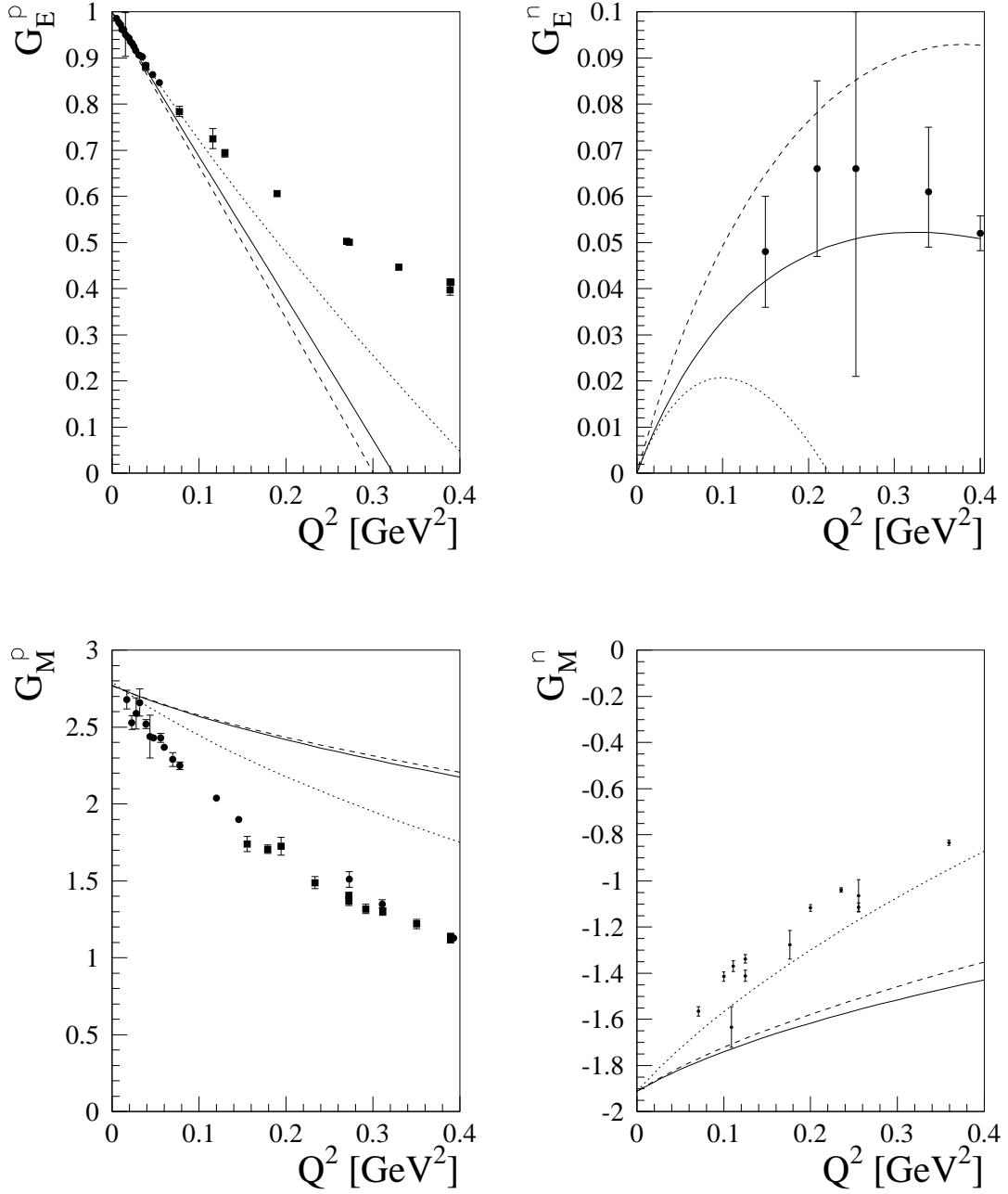, width=14.5truecm} 
\caption{\label{emffnsachsq3pict} 
The Sachs form factors of the nucleon at ${\cal O}(q^3)$. 
The solid, dashed, and dotted lines refer
to the results in the EOMS scheme, the infrared regularization 
\cite{Kubis:2000zd}, and HB$\chi$PT  
\cite{Bernard:1992qa,Fearing:1997dp}, respectively.
   The experimental data for $G_E^p$, $G_E^n$, $G_M^p$, and
$G_M^n$ are taken from Refs.\
\cite{Price:zk,Berger:1971kr,Hanson:vf,Simon:hu},
\cite{Eden:ji,Passchier:1999cj,Ostrick:xa,Herberg:ud,Becker:tw},
\cite{Berger:1971kr,Hanson:vf,Janssens,Hohler:1976ax}, and
\cite{Markowitz:hx,Anklin:ae1,Bruins:ns,Anklin:ae2,Xu:2000xw,Kubon:2001rj}, 
respectively.
} 
\end{figure}

   In order to discuss the Sachs form factors resulting from an 
${\cal O}(q^3)$ calculation, we need to recalculate the adapted values for 
the LECs (see Table \ref{lecwerte2})
since the values of Table \ref{lecwerte} were obtained from an 
${\cal O}(q^4)$ calculation.
   Note that the changes in the LECs turn out to be reasonably small. 
\begin{table}[ht] 
\begin{center} 
\begin{tabular}{|c||c|c|c|c|c|c|} 
  \hline 
     & $c_6$ & $c_7$ & $d_6$ & $d_7$\\ 
   \hline\hline 
   EOMS & 1.34 & -0.15 & -0.70 & -0.49\\ 
   IR & 1.36 & -0.23 & 0.50 & -0.73\\ \hline 
\end{tabular} 
\end{center} 
\caption{\label{lecwerte2} 
   Values for the relevant low-energy constants in the extended on-mass-shell 
renormalization scheme (EOMS) and in the infrared regularization scheme (IR) 
within an ${\cal O}(q^3)$ calculation of relativistic chiral perturbation 
theory. 
   The LECs $c_i$ are given in units of GeV$^{-1}$ and the $d_i$ in units of 
GeV$^{-2}$.} 
\end{table} 
   Figure \ref{emffnsachsq3pict} shows our results for the Sachs form factors 
at ${\cal O}(q^3)$ together with experimental data as well as the 
infrared regularization and HB$\chi$PT results at ${\cal O}(q^3)$.
   For both Sachs form factors of the proton and the magnetic 
Sachs form factor of the neutron the two relativistic calculations 
are almost identical while the HB$\chi$PT results clearly deviate from the 
relativistic calculations.
   This difference may be interpreted as being due to relativistic 
corrections which, in HB$\chi$PT, would have to show up in higher orders.
   It is somewhat surprising that in all three cases the HB$\chi$PT results
are closer to the experimental data.
   The slopes of the electric form factors are essentially determined
by fitting the mean square electric radii.
   On the other hand, at ${\cal O}(q^3)$ the magnetic radii do not
contain any free parameters. 
   This explains the difference in the behavior of the slopes of the magnetic 
form factors between the relativistic calculation on the one hand and the 
HB$\chi$PT calculation on the other hand, originating from the fact that 
the relativistic calculations at ${\cal O}(q^3)$ for the magnetic
radii also contain higher-order terms in $M$, whereas 
HB$\chi$PT only involves the {\em leading}-order term.
   Finally, the electric form factor of the neutron is very small and all 
three theoretical curves disagree. 
   Here, the relativistic ${\cal O}(q^3)$ calculations are in better
agreement with the experimental data.

\begin{figure}
\epsfig{file=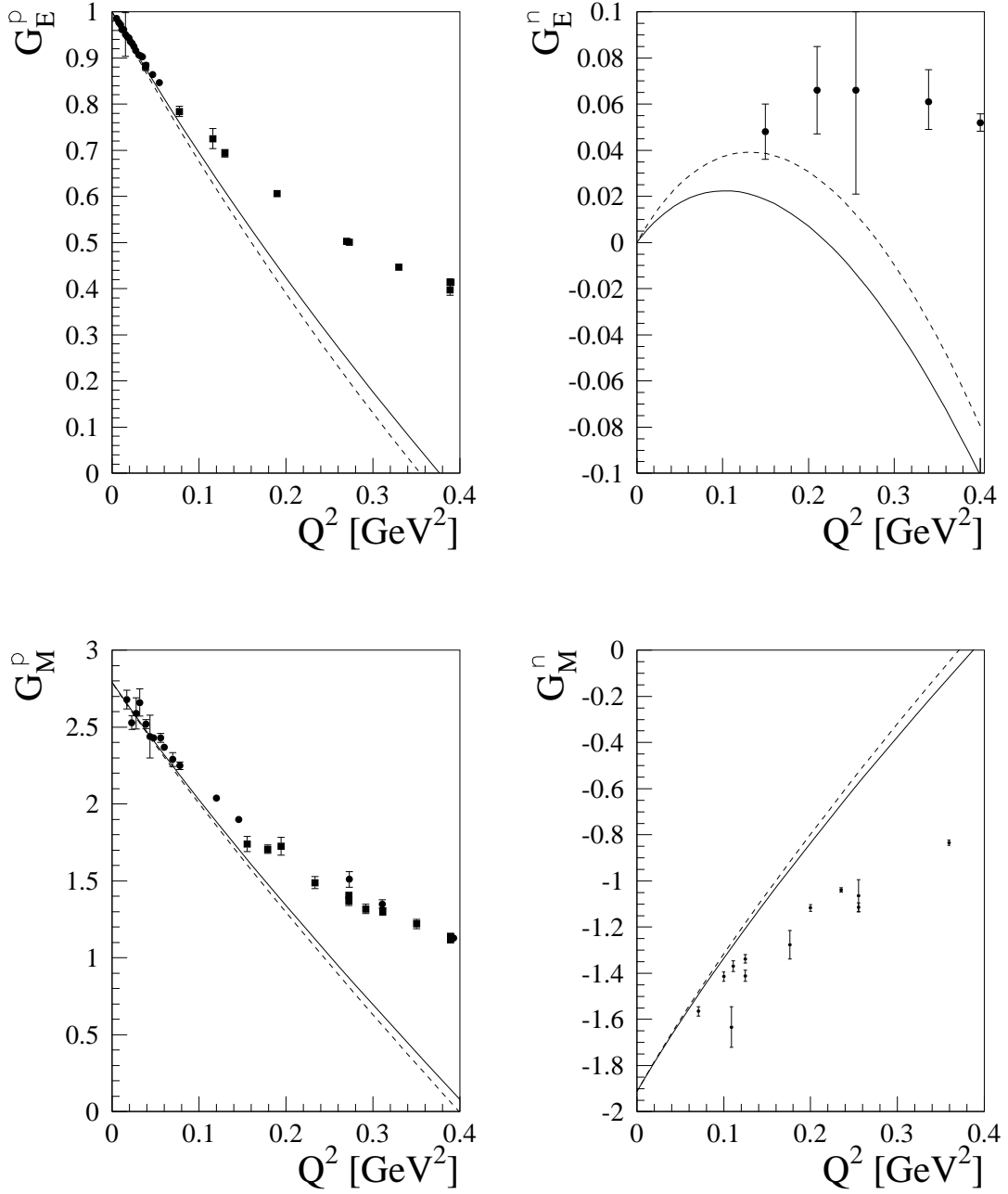, width=14.5truecm} 
\caption{\label{emffnsachspict}  
The Sachs form factors of the nucleon at ${\cal O}(q^4)$. 
The solid and dashed lines refer to the results in the EOMS scheme and
the infrared regularization \cite{Kubis:2000zd}, respectively.
   The experimental data for $G_E^p$, $G_E^n$, $G_M^p$, and
$G_M^n$ are taken from Refs.\
\cite{Price:zk,Berger:1971kr,Hanson:vf,Simon:hu},
\cite{Eden:ji,Passchier:1999cj,Ostrick:xa,Herberg:ud,Becker:tw},
\cite{Berger:1971kr,Hanson:vf,Janssens,Hohler:1976ax}, and
\cite{Markowitz:hx,Anklin:ae1,Bruins:ns,Anklin:ae2,Xu:2000xw,Kubon:2001rj}, 
respectively.}
\end{figure}

\begin{figure}
\epsfig{file=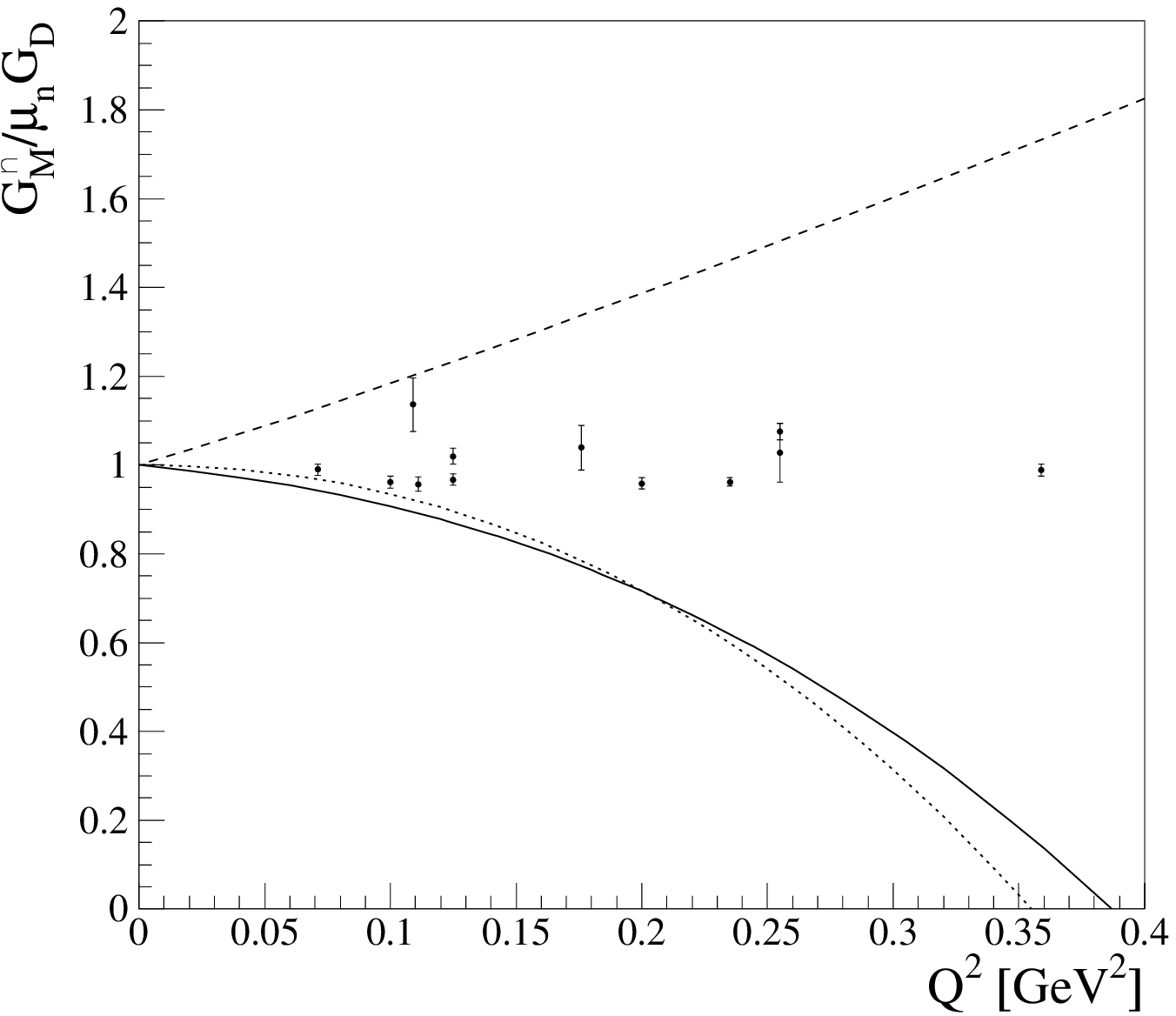, height=9truecm} 
\caption{\label{gmffdipoln} 
The magnetic form factor of the neutron divided by $\mu_n G_D$. 
The solid and dashed lines refer to the results in the EOMS scheme
at ${\cal O}(q^4)$ and ${\cal O}(q^3)$, respectively.
The dotted line is the linear approximation of the dipole form factor, 
i.e., $[1-2Q^2/(0.71\mbox{GeV}^2)]/G_D(Q^2)$. 
The data are taken from Refs.\
\cite{Markowitz:hx,Anklin:ae1,Bruins:ns,Anklin:ae2,Xu:2000xw,Kubon:2001rj}.}
\end{figure}

\begin{figure}
\epsfig{file=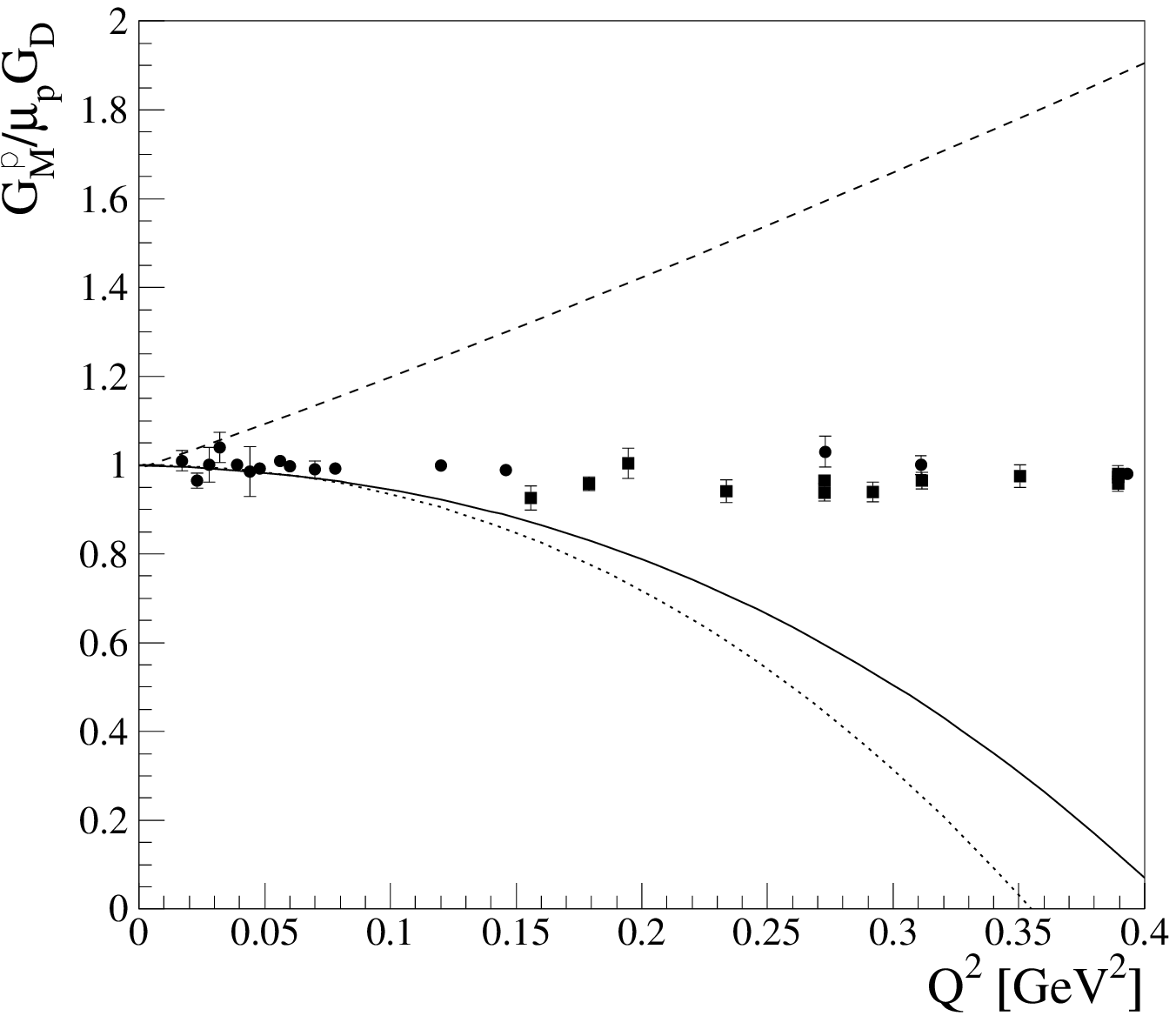, height=9truecm} 
\caption[]{\label{gmffdipolp} 
The magnetic form factor of the proton divided by $\mu_p G_D$. 
The solid and dashed lines refer to the results in the EOMS scheme
at ${\cal O}(q^4)$ and ${\cal O}(q^3)$, respectively.
The dotted line is the linear approximation of the dipole form factor, 
i.e., $[1-2Q^2/(0.71\mbox{GeV}^2)]/G_D(Q^2)$.
The data are taken from Refs.\ 
\cite{Berger:1971kr,Hanson:vf,Janssens,Hohler:1976ax}.
} 
\end{figure}

\begin{figure}
\epsfig{file=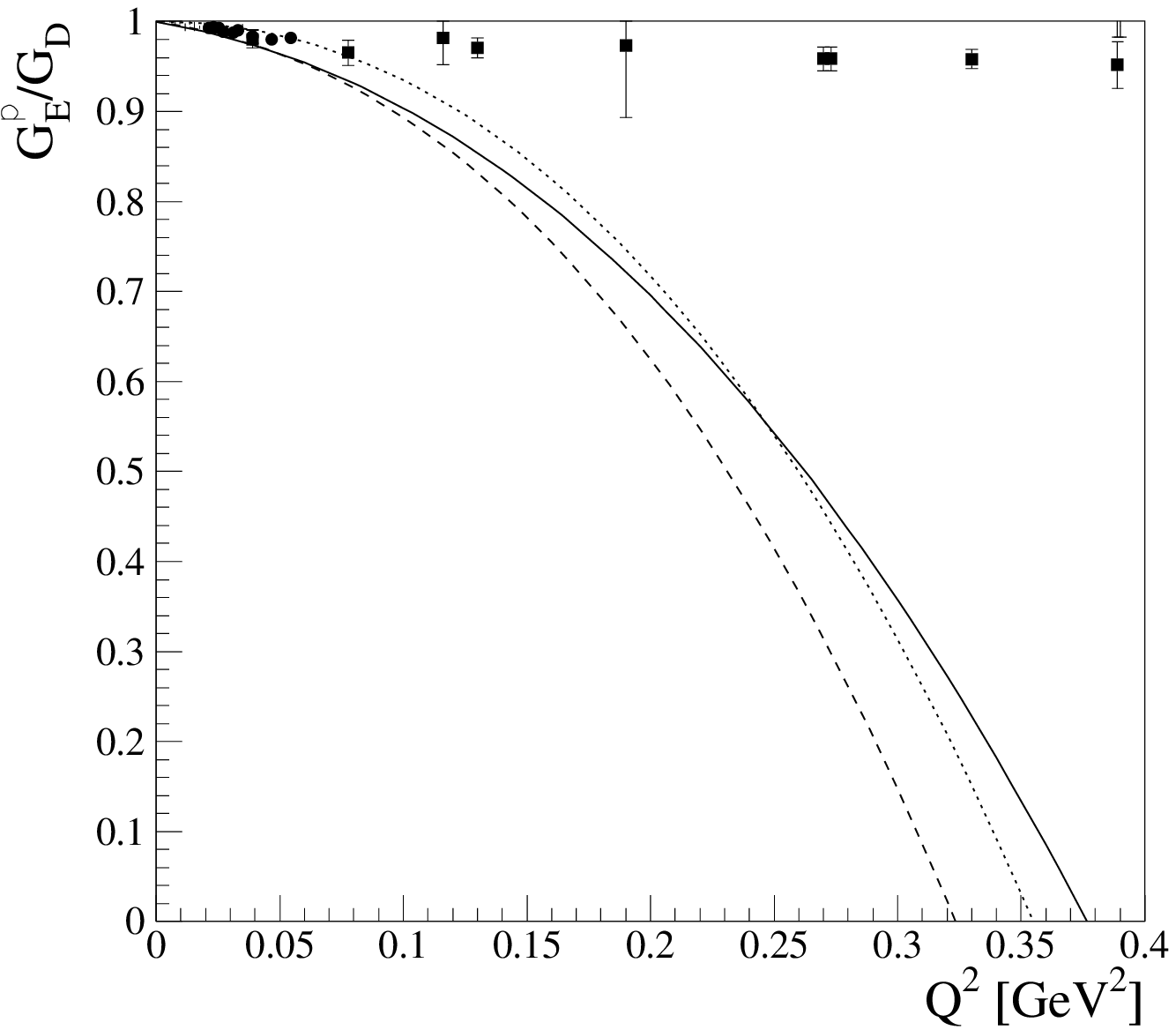, height=9truecm} 
\caption[]{\label{geffdipolp} 
The electric form factor of the proton divided by $G_D$. 
The solid and dashed lines refer to the results in the EOMS scheme
at ${\cal O}(q^4)$ and ${\cal O}(q^3)$, respectively.
The dotted line is the linear approximation of the dipole form factor, 
i.e., $[1-2Q^2/(0.71\mbox{GeV}^2)]/G_D(Q^2)$.
The data are taken from Refs.\ 
\cite{Price:zk,Berger:1971kr,Hanson:vf,Simon:hu}.
}
\end{figure}

   Figure \ref{emffnsachspict} shows the results of the two relativistic
calculations for the Sachs form factors at ${\cal O}(q^4)$.
   The description of $G_E^p$, $G_M^p$, and $G_M^n$ is only marginally
better than that of the ${\cal O}(q^3)$ calculation.
   For the very-small $Q^2$ region this is due to additional free parameters
which have been adjusted to the magnetic radii.
   Clearly, the ${\cal O}(q^4)$ results do not generate sufficient curvature
which is emphasized in Figs.\ \ref{gmffdipoln}, 
\ref{gmffdipolp}, and \ref{geffdipolp}, showing the form factors and data
divided by the dipole form factor $G_D$ and normalized to one at the
origin.
   Finally, the situation for $G_E^n$ seems to be even worse, where we find
better agreement with the experimental data for the ${\cal O}(q^3)$ results.
   We conclude that the perturbation series converges, at best, slowly and
that higher-order contributions must play an important role.

\section{\label{summary}Summary} 
   We have calculated the electromagnetic form factors of the nucleon 
at ${\cal O}(q^4)$ (one-loop order) in relativistic chiral perturbation 
theory using the extended on-mass-shell renormalization scheme of
Ref.\ \cite{Fuchs:2003qc}.
   The relevant low-energy coupling constants have been fitted to
the anomalous magnetic moments and to the charge and magnetic mean
square radii.
   As a general trend the results for both the proton form factors
$G_E^p$ and $G_M^p$ and the magnetic neutron form factor $G_M^n$
do not show sufficient curvature to generate agreement with the
experimental data beyond small values of  
$Q^2\sim 0.1\ \mbox{GeV}^2$.
   Our results are very similar to those of the infrared regularization 
\cite{Kubis:2000zd} and the small differences between the two methods
are related to the way how the regular higher-order terms of loop integrals
are treated.
   In the case of the electric neutron form factor $G_E^n$, in both
renormalization schemes the description does not seem to improve
when going from ${\cal O}(q^3)$ to ${\cal O}(q^4)$.
   We conclude that a relativistic treatment at the one-loop level
using nucleon and pion degrees of freedom only is not sufficient to describe
the form factors for $Q^2\geq 0.1\ \mbox{GeV}^2$ and
that higher-order contributions must play an important role. 
   Such higher-order contributions either have to be studied in the framework
of a two-loop calculation or by explicitly including other dynamical
degrees of freedom such as vector mesons \cite{Kubis:2000zd}.
   We stress that the EOMS renormalization scheme allows for also
implementing a consistent power counting in relativistic baryon
chiral perturbation theory when vector (and axial-vector) mesons
are explicitly included \cite{Fuchs}.

\acknowledgments 
   The work of T.F.~and S.S.~was supported by the Deutsche 
Forschungsgemeinschaft (SFB 443).
   J.G.~acknowledges the support of the Alexander von Humboldt Foundation.

\appendix 
\section{Loop integrals} 
\subsection{Definition of the loop integrals} 
In the appendix we use the following notation 
$$ P^\mu=p^\mu_i + p^\mu_f, \qquad q^\mu=p_f^\mu - p_i^\mu, \qquad 
  t = q^2. $$ 
The loop integrals with one and two internal lines are defined as follows: 
\begin{displaymath} 
  I_\pi = i\int \frac{d^nk}{(2\pi)^n} \frac{1}{k^2-M^2+i\epsilon},
\end{displaymath}
\begin{displaymath}
  I_N = i\int \frac{d^nk}{(2\pi)^n} \frac{1}{k^2-m^2+i\epsilon},
\end{displaymath}
\begin{displaymath} 
  I_{\pi\pi}(t) = i \int \frac{d^nk}{(2\pi)^n}
    \frac{1}{[k^2-M^2 + i\epsilon][(k+q)^2 - M^2 + i\epsilon]},
\end{displaymath} 
\begin{displaymath}
  q^\mu I_{\pi\pi}^{(q)}(t) = i \int \frac{d^nk}{(2\pi)^n}
    \frac{k^\mu}{[k^2-M^2 + i\epsilon][(k+q)^2 - M^2 + i\epsilon]},
\end{displaymath} 
\begin{displaymath}
  t\,g^{\mu\nu}I^{(00)}_{\pi\pi}(t) + q^\mu q^\nu I^{(qq)}_{\pi\pi}(t) 
    = i  \int \frac{d^nk}{(2\pi)^n}
  \frac{k^\mu k^\nu}{[k^2-M^2 + i\epsilon][(k+q)^2 - M^2+i\epsilon]},
\end{displaymath}
\begin{displaymath} 
  I_{NN}(t) = i \int \frac{d^nk}{(2\pi)^n}
    \frac{1}{[k^2-m^2 + i\epsilon][(k+q)^2 - m^2 + i\epsilon]},
\end{displaymath}
\begin{displaymath} 
  I_{\pi N}(p^2) = i \int \frac{d^nk}{(2\pi)^n}
    \frac{1}{[k^2-M^2 + i\epsilon][(k-p)^2 - m^2 + i\epsilon]},
\end{displaymath}
\begin{displaymath} 
  p^\mu I^{(p)}_{\pi N}(p^2) = i \int \frac{d^nk}{(2\pi)^n}
    \frac{k^\mu}{[k^2-M^2 + i\epsilon][(k-p)^2 - m^2+i\epsilon]}. 
\end{displaymath} 
   In the following loop integrals involving three internal lines we always
assume on-shell kinematics, $p_i^2=p_f^2=m_N^2$,
\begin{displaymath} 
   I_{\pi\pi N}(t) = i \int \frac{d^nk}{(2\pi)^n}
       \frac{1}{\left[k^2-M^2+i\epsilon\right] \left[(k+q)^2-M^2 
       +i\epsilon\right][(k-p_i)^2-m^2+i\epsilon]},
\end{displaymath}
\begin{displaymath}
 P^\mu I^{(P)}_{\pi \pi N}(t) - \frac{1}{2}q^\mu I_{\pi \pi N}(t) = 
    i \int \frac{d^nk}{(2\pi)^n} \frac{k^\mu}{[k^2-M^2 + i\epsilon]
    [(k+q)^2 - M^2 + i\epsilon][(k-p_i)^2 - m^2 + i\epsilon]}, 
\end{displaymath}
\begin{eqnarray*} 
  \lefteqn{ g^{\mu\nu} I_{\pi \pi N}^{(00)}(t) 
  + P^\mu P^\nu I^{(PP)}_{\pi \pi N}(t) 
     + q^\mu q^\nu I^{(qq)}_{\pi \pi N}(t) - \frac{q^\mu P^\nu 
     + P^\mu q^\nu}{2} I^{(P)}_{\pi \pi N}(t)}\\ 
&=&i \int \frac{d^nk}{(2\pi)^n} \frac{k^\mu k^\nu}{[k^2-M^2 + i\epsilon]
[(k+q)^2 - M^2 + i\epsilon][(k-p_i)^2 - m^2 + i\epsilon]},
\end{eqnarray*}
\begin{displaymath}
  I_{\pi N N}(t) = i \int \frac{d^nk}{(2\pi)^n} 
    \frac{1}{[k^2-M^2+i\epsilon][(k-p_i)^2-m^2 + i\epsilon] 
    [(k-p_f)^2 - m^2 + i\epsilon]},
\end{displaymath} 
\begin{displaymath}
   P^\mu I^{(P)}_{\pi N N}(t)=i \int \frac{d^nk}{(2\pi)^n} 
    \frac{k^\mu}{[k^2-M^2 + i\epsilon] 
    [(k-p_i)^2 - m^2 + i\epsilon][(k-p_f)^2 - m^2 + i\epsilon]}, 
\end{displaymath} 
\begin{eqnarray*} 
\lefteqn{g^{\mu\nu} I_{\pi N N}^{(00)}(t) + P^\mu P^\nu I^{(PP)}_{\pi N N}(t) 
     + q^\mu q^\nu I^{(qq)}_{\pi N N}(t)}\\
&=&i \int \frac{d^nk}{(2\pi)^n} \frac{k^\mu k^\nu} 
     {[k^2-M^2 + i\epsilon][(k-p_i)^2 - m^2 + i\epsilon][(k-p_f)^2 
     - m^2 + i\epsilon]}. 
\end{eqnarray*} 

\subsection{Scalar loop integrals} 
   Defining 
\begin{displaymath} 
\bar\lambda ={m^{n-4}\over 16\pi^2}\left\{ {1\over n-4}-{1\over 2} 
\left[ \ln (4\pi) +\Gamma '(1)+1\right]\right\}, 
\end{displaymath} 
and 
\begin{displaymath} 
\Omega=\frac{p^2-m^2-M^2}{2mM},
\end{displaymath}
the scalar loop integrals are given by \cite{Fuchs:2003qc}
\begin{displaymath}
I_\pi=2M^2\bar{\lambda}+\frac{M^2}{8\pi^2}\ln\left(\frac{M}{m}\right), 
\end{displaymath}
\begin{displaymath}
I_N=2m^2\bar{\lambda},
\end{displaymath}
\begin{displaymath}
I_{\pi\pi}(t)=2\bar{\lambda}+\frac{1}{16\pi^2}\left[
1 + 2\ln\left(\frac{M}{m}\right) 
                    + J^{(0)}\left(\frac{t}{M^2}\right)\right],
\end{displaymath}
\begin{displaymath}
I_{NN}(t)=2\bar{\lambda}+\frac{1}{16\pi^2}\left[1+J^{(0)}
\left(\frac{t}{m^2}\right)\right]
\end{displaymath}
\begin{displaymath}
I_{\pi N}(p^2)=2\bar{\lambda}+\frac{1}{16\pi^2}\left[-1
+\frac{p^2-m^2+M^2}{p^2}\ln\left(\frac{M}{m}\right)
+\frac{2mM}{p^2}F(\Omega)\right],
\end{displaymath}
where 
\begin{eqnarray*} 
J^{(0)}(x) 
&=&\int_0^1 dz \ln[1+x(z^2-z)-i\epsilon]\\ 
&=& 
 \left \{ \begin{array}{ll} 
-2-\sigma\ln\left(\frac{\sigma-1}{\sigma+1}\right),&x<0,\\ 
-2+2\sqrt{\frac{4}{x}-1}\,\mbox{arccot} 
\left(\sqrt{\frac{4}{x}-1}\right),&0\le x<4,\\ 
-2-\sigma\ln\left(\frac{1-\sigma}{1+\sigma}\right)-i\pi\sigma,& 
4<x, 
\end{array} \right. 
\end{eqnarray*} 
with 
\begin{displaymath} 
\sigma(x)=\sqrt{1-\frac{4}{x}},\quad x\notin [0,4], 
\end{displaymath} 
and 
\begin{eqnarray*} 
F(\Omega) 
&=& 
\left \{ \begin{array}{ll} 
\sqrt{\Omega^2-1}\ln\left(-\Omega-\sqrt{\Omega^2-1}\right),&\Omega\leq -1,\\
\sqrt{1-\Omega^2}\arccos(-\Omega),&-1\leq\Omega\leq 1,\\
\sqrt{\Omega^2-1}\ln\left(\Omega+\sqrt{\Omega^2-1}\right)
-i\pi\sqrt{\Omega^2-1},&1\leq
\Omega.
\end{array} \right. 
\end{eqnarray*} 
   In our numerical analysis, we made use of a parameterization in terms
of a Feynman parameter integral to analyze the loop integrals with three 
internal lines.
   Such integrals may also be determined using a dispersion relation
(see App.~C of Ref.\ \cite{Fuchs:2003qc}).

\subsection{Reduction of the tensorial loop integrals} 
   The reduction of the tensorial loop integrals to the corresponding scalar 
ones can be performed in the standard way \cite{Passarino:1978jh}. 
   We obtain the following results: 
\begin{displaymath}
I_{\pi\pi}^{(q)}(t) = -\frac{1}{2}I_{\pi\pi}(t),
\end{displaymath}
\begin{eqnarray*}
I_{\pi\pi}^{(qq)}(t)  &=&  \frac{1}{(n-1)t} \left[\frac{n-2}{2}
I_\pi + \left(\frac{n}{4}\,t - M^2\right)I_{\pi\pi}(t)\right] \\ 
&=& \frac{1}{t}\left[\frac{1}{3}I_\pi + 
       \frac{1}{3}\left(t-M^2\right)I_{\pi\pi}(t) 
       + \frac{1}{144\pi^2}\left(3M^2-\frac{t}{2}\right)\right],
\end{eqnarray*}
\begin{eqnarray*}
  I_{\pi\pi}^{(00)}(t) &=& \frac{1}{(n-1)t}\left[\frac{1}{2}I_\pi + 
      \frac{1}{4}(4M^2-t)I_{\pi\pi}(t) \right]\\ &=& 
       \frac{1}{t}\left[\frac{1}{6}I_\pi(t) + 
       \frac{1}{12}\left(4M^2-t\right)I_{\pi\pi}(t) 
     -\frac{1}{8\pi^2}\left(\frac{M^2}{6} - \frac{t}{36}\right)\right],\\ 
\end{eqnarray*}
\begin{displaymath}
I_{\pi N}^{(p)}(p^2) = \frac{1}{2p^2}\left[I_N - I_\pi 
     + (p^2-m^2+M^2)I_{\pi N}(p^2)\right], 
\end{displaymath}
\begin{displaymath}
I_{\pi \pi N}^{(P)}(t) = \frac{1}{2(4m_N^2-t)}\left[ 
 (2M^2-t)I_{\pi \pi N}(t) + 2 I_{\pi N}(m_N^2) - 2 I_{\pi\pi}(t)\right],
\end{displaymath}
\begin{eqnarray*}
I_{\pi \pi N}^{(00)}(t) &=& \frac{1}{4(n-2)}\left[2I_{\pi N}(m_N^2) 
   + (4M^2-t)I_{\pi \pi N}(t) - 2(2M^2-t)I_{\pi \pi N}^{(P)}(t)\right]\\ 
&=& \frac{1}{8}\left[2I_{\pi N}(m_N^2) 
   + (4M^2-t)I_{\pi \pi N}(t) - 2(2M^2-t)I_{\pi \pi N}^{(P)}(t) 
   - \frac{1}{8\pi^2}\right],
\end{eqnarray*}
\begin{eqnarray*}
 I_{\pi \pi N}^{(PP)}(t)&=& \frac{1}{4(n-2)(4m_N^2-t)}\left[-2I_{\pi N}(m_N^2) 
   + 2(n-2)I_{\pi N}^{(p)}(m_N^2) - (4M^2-t)I_{\pi \pi N}(t)\right.\\
 && \left. + 2(n-1)(2M^2-t)I_{\pi \pi N}^{(P)}(t)\right]\\ 
   &=& \frac{1}{8(4m_N^2-t)}\left[-2I_{\pi N}(m_N^2) 
   + 4I_{\pi N}^{(p)}(m_N^2) - (4M^2-t)I_{\pi \pi N}(t)\right.\nn\\ && 
  \left. + 6(2M^2-t)I_{\pi \pi N}^{(P)}(t) + \frac{1}{8\pi^2}\right],\nn\\
\end{eqnarray*}
\begin{eqnarray*}
 I_{\pi \pi N}^{(qq)}(t) &=&\frac{1}{4(n-2)t}\left\{2(n-3)I_{\pi N}(m_N^2) 
   - 2(n-2)I_{\pi N}^{(p)}(m_N^2) \right.\nn\\ &&\left.- \left[4M^2-(n-1)t 
   \right] I_{\pi \pi N}(t) + 2(2M^2-t) I_{\pi \pi N}^{(P)}(t)\right\}\nn\\ 
   &=& \frac{1}{8t}\left[2I_{\pi N}(m_N^2)- 4I_{\pi N}^{(p)}(m_N^2) 
    - \left(4M^2-3t\right) I_{\pi \pi N}(t) 
    + 2(2M^2-t) I_{\pi \pi N}^{(P)}(t) 
    + \frac{1}{8\pi^2} \right],
\end{eqnarray*}
\begin{displaymath}
  I_{\pi N N}^{(P)}(t) = \frac{1}{4m^2-t}\left[M^2I_{\pi NN}(t) 
     -I_{\pi N}(m_N^2) + I_{NN}(t)\right],
\end{displaymath}
\begin{eqnarray*} 
  I_{\pi N N}^{(00)}(t) &=& \frac{1}{n-2}\left\{\left[I_{\pi N N}(t) - 
     I_{\pi N N}^{(P)}(t)\right]M^2 + \frac{1}{2}I_{NN}(t)\right\}\\ 
     &=& \frac{1}{2}\left\{\left[I_{\pi N N}(t) - 
     I_{\pi N N}^{(P)}(t)\right]M^2 + \frac{1}{2}I_{NN}(t) 
     - \frac{1}{32\pi^2}\right\}, 
\end{eqnarray*}
\begin{eqnarray*} 
 I_{\pi N N}^{(PP)}(t) &=& \frac{1}{(n-2)(4m_N^2-t)}\left\{\left[ 
     (n-1) I_{\pi N N}^{(P)}(t) - I_{\pi N N}(t)\right]M^2 
     -\frac{n-2}{2}I^{(p)}_{\pi N}(m_N^2) 
   + \frac{n-3}{2}I_{NN}(t)\right\}\\ 
   &=& \frac{1}{2(4m_N^2-t)}\left\{\left[ 
     3 I_{\pi N N}^{(P)}(t) - I_{\pi N N}(t)\right]M^2 
     - I^{(p)}_{\pi N}(m_N^2) + \frac{1}{2}I_{NN}(t) 
     + \frac{1}{32\pi^2}\right\},
\end{eqnarray*}
\begin{eqnarray*}
  I_{\pi N N}^{(qq)}(t) &=& \frac{1}{(n-2)t}\left\{\left[ I_{\pi N N}^{(P)}(t) 
     - I_{\pi N N}(t)\right]M^2 + \frac{n-2}{2}I^{(p)}_{\pi N}(m_N^2) 
     -\frac{1}{2}I_{NN}(t)\right\}\\ 
    &=& \frac{1}{2t}\left\{\left[ I_{\pi N N}^{(P)}(t) 
     - I_{\pi N N}(t)\right]M^2 + I^{(p)}_{\pi N}(m_N^2) 
     -\frac{1}{2}I_{NN}(t) + \frac{1}{32\pi^2}\right\}. 
\end{eqnarray*}

\section{Contributions to the Dirac and Pauli form factors} 
\label{emffanh} 
\begin{figure}
\epsfig{file=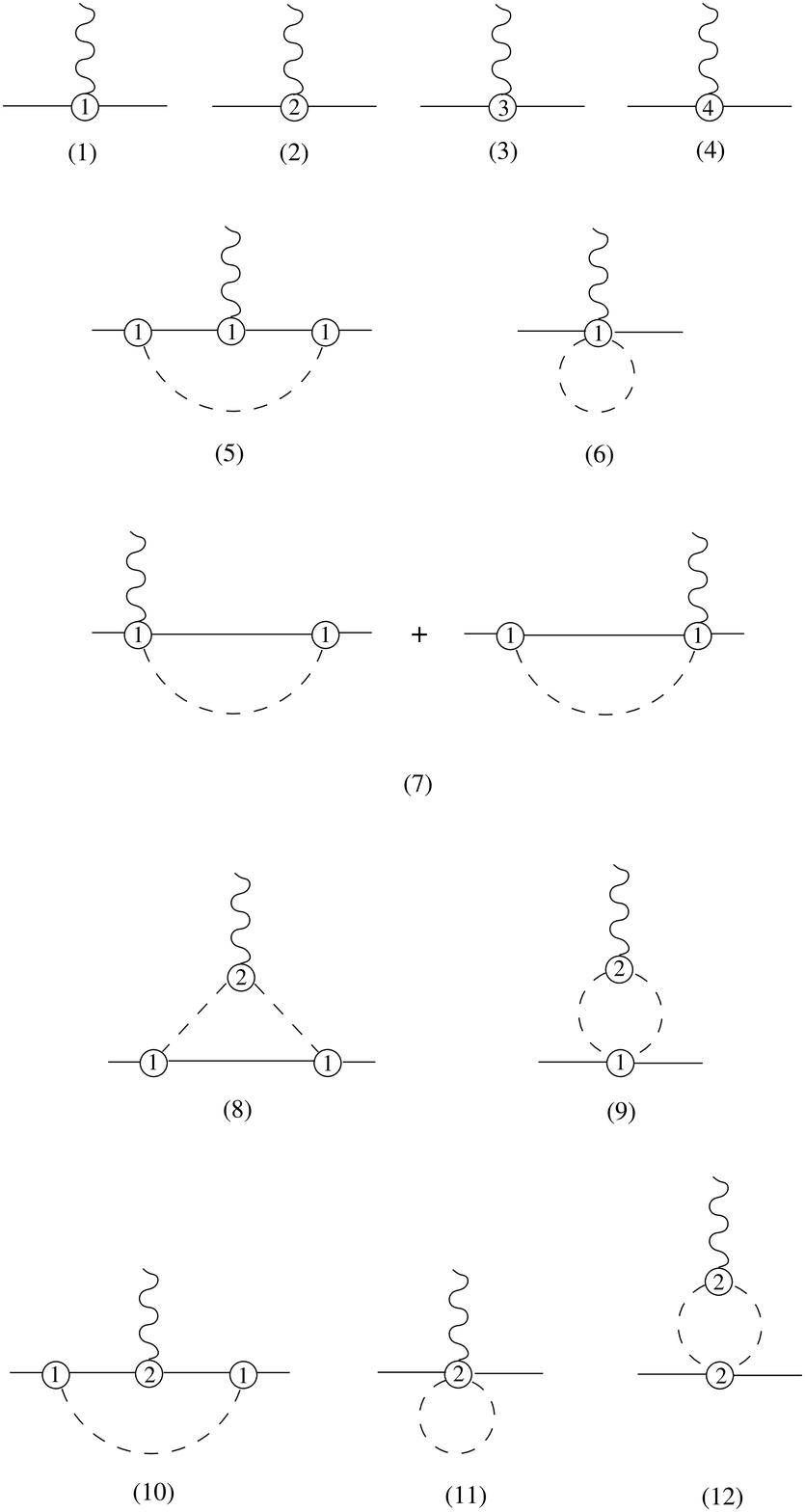,height=20cm} 
\caption{\label{emffgraphen} Feynman diagrams contributing to the 
electromagnetic form factors up to and including ${\cal O}(q^4)$.
   External-leg corrections are not shown.
   Solid, dashed, and wiggly lines refer to nucleons, pions, and photons, 
respectively.} 
\end{figure}

The unrenormalized expressions for the Dirac form factor $F_1$ read 
\begin{displaymath}
F_1^{\,1+3}= \frac{1+\tau_3}{2} - \left( \tau_3\,d_6 + 2d_7 \right)\, t,
\end{displaymath}
\begin{eqnarray*}
F_1^{\,5} &=& \frac{\gA0^2}{8F^2} \, (3-\tau_3) \,
\{I_\pi + 4m_N^2 I^{(p)}_{\pi N}(m_N^2) - 4m_N^2[M^2 I_{\pi NN}(t) 
+I_{NN}(t)]\\
&& + 8m_N^2 I_{\pi NN}^{(00)}(t)+32m_N^4 I_{\pi NN}^{(PP)}(t) \},
\end{eqnarray*}
\begin{displaymath}
F_1^{\,6} = -\frac{\tau_3}{2F^2}\,I_\pi,
\end{displaymath}
\begin{displaymath}
F_1^{\,7} = \frac{\gA0^2}{F^2}\,\tau_3\,[I_\pi+2 m_N^2 I^{(p)}_{\pi N}(m_N^2)],
\end{displaymath}
\begin{displaymath}
F_1^{\,8} = -\frac{\gA0^2}{F^2} \, \tau_3\,[ t I_{\pi\pi}^{(00)}(t) 
    +4m_N^2 I_{\pi\pi N}^{(00)}(t) +16m_N^4 I_{\pi\pi N}^{(PP)}(t)],
\end{displaymath}
\begin{displaymath}
F_1^{\,9}= \frac{\tau_3}{F^2}\,t \,I^{(00)}_{\pi \pi}(t),
\end{displaymath}
\begin{displaymath}
F_1^{\,10} = \frac{2m_N^3\gA0^2}{F^2}\left( 3c_7 - 2 c_6\tau_3 \right) 
    \,t\,I_{\pi NN}^{(PP)}(t),
\end{displaymath}
\begin{displaymath}
F_1^{\,11} = \frac{3M^2}{4m_N F^2}\,(1+\tau_3)\,c_2\,\left(I_\pi 
  - \frac{M^2}{32\pi^2} \right)={\cal O}(q^4).
\end{displaymath} 
   The contributions to the Pauli form factor $F_2$ are given by 
\begin{eqnarray*} 
F_2^{\,2+3+4} &=& 2m_N\tau_3\,c_6 + m_N c_7 
    + \left( \tau_3\,d_6 + 2d_7 
       \right)\, t + 2m_N ( 2\,e_{54} + \tau_3\,e_{74} )\, t \\ 
    && - 8m_N M^2 \left( 2\,e_{105}+\tau_3\,e_{106} \right),
\end{eqnarray*}
\begin{displaymath}
F_2^{\,5}= -\frac{\gA0^2}{F^2} \, (3-\tau_3) \, 4 m_N^4 \, 
    I_{\pi NN}^{(PP)}(t), 
\end{displaymath}
\begin{displaymath}
F_2^{\,8} = \frac{\gA0^2}{F^2}\,\tau_3 \, 
    16m_N^4I_{\pi\pi N}^{(PP)}(t), 
\end{displaymath}
\begin{eqnarray*}
F_2^{\,10}&=& -\frac{\gA0^2m_N}{4F^2}\left( 3c_7 - 2 c_6\tau_3\right) 
    \{ I_\pi + 4m_N^2 I^{(p)}_{\pi N}(m_N^2) + 4m_N^2M^2 I_{\pi NN}(t)\\ 
&& +4m_N^2 I_{NN}(t)-16m_N^2I_{\pi NN}^{(00)}(t) 
    + 8m_N^2\,t\,[ I_{\pi NN}^{(PP)}(t)-I_{\pi NN}^{(qq)}(t)]\}, 
\end{eqnarray*}
\begin{displaymath}
F_2^{\,11} =  -\frac{3M^2}{4m_NF^2}\,\left(1+\tau_3\right)\,c_2\, 
     \left( I_\pi - \frac{M^2}{32\pi^2} \right) 
     - 2\frac{m_N}{F^2} \tau_3\,c_6\, I_\pi,
\end{displaymath} 
\begin{displaymath}
F_2^{\,12} = \frac{4m_N}{F^2} \,\tau_3 \, c_4 \, t\, 
      I_{\pi\pi}^{(00)}(t).
\end{displaymath}
   The upper indices refer to the corresponding diagrams of Fig.\ 
\ref{emffgraphen}. 
   Note that we did not distinguish between $m$ and $m_N$, because the 
difference only shows up at higher orders.
   Our results agree with the expressions of Ref.\ \cite{Kubis:2000zd} 
up to terms vanishing in infrared regularization.


\end{document}